\documentclass[a4paper,conference]{IEEEtran}
\usepackage{hyperref}  
\IEEEoverridecommandlockouts
\usepackage{cite}
\usepackage{amsmath,amssymb,amsfonts}
\usepackage{algorithmic}
\usepackage{graphicx}
\usepackage{textcomp}
\usepackage{xcolor}
\usepackage{adjustbox} 
\usepackage{multicol} 
\def\BibTeX{{\rm B\kern-.05em{\sc i\kern-.025em b}\kern-.08em
    T\kern-.1667em\lower.7ex\hbox{E}\kern-.125emX}}

 \usepackage[pscoord]{eso-pic}

\begin{document}

\title{Optimal Scalogram for Computational Complexity Reduction in Acoustic Recognition Using Deep Learning\\
}

\author{
\IEEEauthorblockN{Dang Thoai Phan\IEEEauthorrefmark{1}, Tuan Anh Huynh\IEEEauthorrefmark{2}, Van Tuan Pham\IEEEauthorrefmark{3},\\
Cao Minh Tran\IEEEauthorrefmark{4}, Van Thuan Mai\IEEEauthorrefmark{5}, Ngoc Quy Tran\IEEEauthorrefmark{6}}
\IEEEauthorblockA{\IEEEauthorrefmark{1}BHT University of Applied Sciences and Technology, Berlin, Germany; thoai.phandang@gmail.com}
\IEEEauthorblockA{\IEEEauthorrefmark{2}Faculty of Software Engineering, University of Information Technology, Ho Chi Minh city, Vietnam\\
Vietnam National University, Ho Chi Minh city, Vietnam; anhht@uit.edu.vn}
\IEEEauthorblockA{\IEEEauthorrefmark{3}Artificial Intelligence, Yokogawa Votiva Solutions, Ho Chi Minh city, Vietnam; tuan.pham@votivasoft.com}
\IEEEauthorblockA{\IEEEauthorrefmark{4}Information Technology, Nguyen Tat Thanh University, Ho Chi Minh city, Vietnam; trancaominhkg@gmail.com}
\IEEEauthorblockA{\IEEEauthorrefmark{5}Smart Ocean Mobility, Changwon National University, Changwon, Republic of Korea; maivanthuan996@gmail.com}
\IEEEauthorblockA{\IEEEauthorrefmark{6}Software Engineering, FPT University Hanoi,
Hanoi, Vietnam; quytnhe161643@fpt.edu.vn}
}

\maketitle

\begin{abstract}
The Continuous Wavelet Transform (CWT) is an effective tool for feature extraction in acoustic recognition using Convolutional Neural Networks (CNNs), particularly when applied to non-stationary audio. However, its high computational cost poses a significant challenge, often leading researchers to prefer alternative methods such as the Short-Time Fourier Transform (STFT). To address this issue, this paper proposes a method to reduce the computational complexity of CWT by optimizing the length of the wavelet kernel and the hop size of the output scalogram. Experimental results demonstrate that the proposed approach significantly reduces computational cost while maintaining the robust performance of the trained model in acoustic recognition tasks.
\end{abstract}

\section{Introduction}

\begin{figure}[t]
\centering
\includegraphics[width=0.4\textwidth]{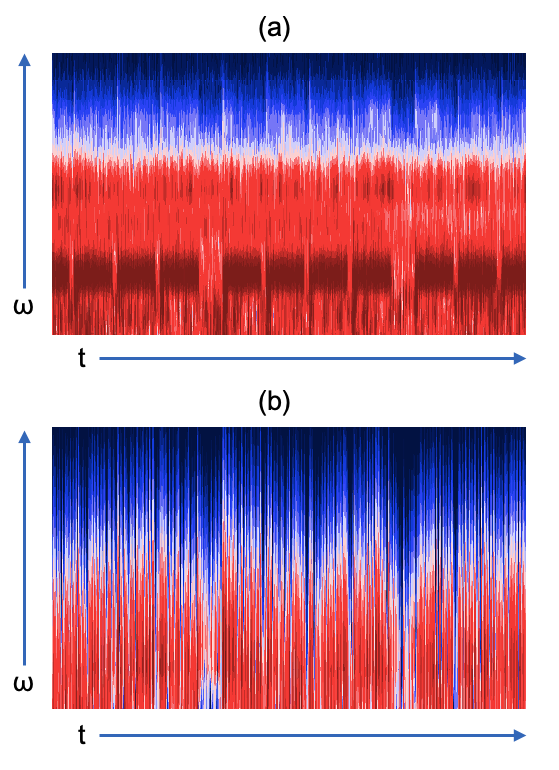}
\caption{Scalogram of CWT (a); optCWT (b)}
\label{Scalograms}
\end{figure}

Feature extraction from time-series signals is a critical step in many deep learning applications, particularly when combined with architectures such as Convolutional Neural Networks (CNNs) \cite{jin2025wave, 10.1007/978-3-031-84457-7_41,trappolini2025quantized, cao2025prediction, huynh2022skin}. Among the various techniques available, the Continuous Wavelet Transform (CWT) has emerged as a widely adopted method, especially in processing acoustic signals, due to its multiresolution analysis capability \cite{guo2022review} that enhances model performance \cite{10797444, 10.1007/978-3-031-84457-7_41}. Significant research efforts have been dedicated to leveraging CWT for improved feature representation \cite{ZHAO2024464582, shabber2025scalogram, negi2025scalogram, yousefpour2025bridge, el2025enhancing, lee2025prediction}.

For instance, Copiaco \cite{copiaco2019scalogram} utilized scalograms derived from the CWT as spectro-temporal features for domestic audio classification. These scalograms were input into a hybrid model comprising CNNs and a Support Vector Machine (SVM), resulting in notable performance improvements compared to top-performing baseline models. Similarly, Gupta \cite{gupta2022morlet} employed CNNs with CWT-based scalograms for voice liveness detection. Their method effectively differentiated between genuine and spoofed speech using a handcrafted Morlet wavelet—substantially outperforming the Short-Time Fourier Transform (STFT) spectrogram. In another study, Chatterjee \ \cite{chatterjee2023deep} addressed musical instrument identification by transforming audio samples into CWT-based scalograms and utilizing a combination of Convolutional Siamese Networks and Residual Siamese Networks. Their approach attained an impressive classification accuracy using only five training samples per class from public datasets. Phan \cite{10.1007/978-3-031-84457-7_41} conducted a comparative analysis of the CWT and STFT as inputs to CNN models for non-stationary machine noise classification. While the results indicated superior recognition performance with the CWT, the study also highlighted its substantial computational cost. In subsequent work, Phan \cite{10797444} proposed a down-sampling approach that applies a hop size to the signals prior to the wavelet transform; however, this method introduces aliasing and attenuates high-frequency features.

These recent contributions highlight the effectiveness of CWT as a time-frequency feature extractor \cite{rahula2025arrhythmia, chanu2025pcg, borzelli2025pooled, dades2025quadratic, choudhury2025explainable, vigneshwaran2025stacking,ramteke2025acute, roozbehi2025enhanced, pal2025detection, altay2025cascade, mostafavi2025correction}. Nonetheless, a major limitation lies in its computational burden: CWT must be computed continuously over each sample in a discrete signal, resulting in substantial data volume and potential redundancy due to similarities among adjacent samples. Consequently, alternative methods such as the STFT \cite{kong2024multi, telmem2025impact, ma2025must,liu2025esernet,banerjee2025deep,dua2025melcochleagram, jin2025wave, mekahlia2025comparative, chen2025wavespect} are often preferred for their computational efficiency.

Given this context, there is a growing need for approaches that preserve the advantages of CWT's multiresolution analysis while mitigating its computational demands. Developing such techniques could significantly improve the practicality and scalability of CWT-based models in real-world applications.

\section{Theoretical foundation}

\subsection{Wavelet transform}

WT is a technique that decomposes a signal into a form that better represents the original signal's features for further processing \cite{addison2017illustrated}. In acoustic recognition, WT converts a one-dimensional (1D) time signal into a two-dimensional (2D) time-frequency plane, as described by the calculation formula in (\ref{CWT equation}). WT is a function of time translation \textit{b} and frequency shift \textit{a}. The signal's energy is normalized by the factor 1/\begin{math}\sqrt{\textit{a}}\end{math} to ensure consistent energy levels across all frequency scales. The wavelet is contracted and dilated according to the varying scale, and each scaled wavelet is then shifted along the time axis to convolve with the signal \textit{x(t)}.
\begin{equation}
X _{WT}(\textit{b}, \textit{a}) = \frac{1}{\sqrt{\textit{a}}}\int_{-\infty}^{\infty} \,x(t)\psi^*(\frac{t-\textit{b}}{\textit{a}})\,dt
\label{CWT equation}
\end{equation}
The CWT in discrete form is given by:
\begin{equation}
W_{WT}(b, a) = \frac{1}{\sqrt{\textit{a}}}\sum_{n=0}^{N-1} x[n] \cdot \psi^* \left( \frac{n - b}{a} \right)
\label{CWT discrete form}
\end{equation}
where: \( n \) is the discrete time index, \( x[n] \) is the discrete signal of length \( N \).
CWT for a time-discrete signal is computed by the discrete summation of the dot product within the sampling interval. The translation parameter \textit{b} and scale parameter \textit{a} are in continuous forms, where translation is sample-wise, and scale spans a range of continuous natural numbers. This process produces a coefficient matrix of size (\textit{N}, \textit{a}), where \textit{N} is the data length and \textit{a} is the scales range. The scalogram of CWT is illustrated in Fig. \ref{Scalograms}(a), with the vertical direction representing the frequency/scale (\begin{math}\omega\end{math}/\textit{a}) and the horizontal direction representing the time/translation (\textit{t}/\textit{b}).

The CWT is commonly implemented using the PyWavelets \cite{lee2019pywavelets} library, specifically through the pywt.cwt function. However, this function does not provide options for adjusting the wavelet kernel length or customizing the scalogram size based on specific requirements. This limitation results in a high computational cost, particularly for tasks that require low temporal resolution. Therefore, an approach that optimizes computational efficiency while preserving essential time-frequency features is highly desirable.

\begin{figure*}[t]
\centering
\includegraphics[width=0.8\textwidth]{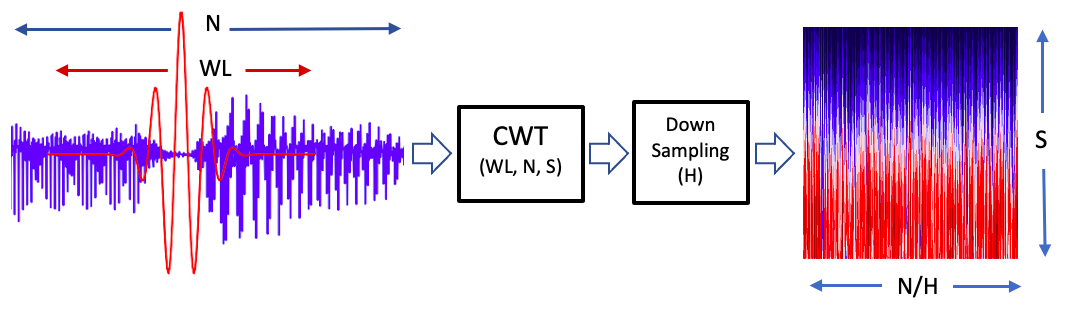}
\caption{Scalogram generation}
\label{ScalogramGeneration}
\end{figure*}

\subsection{Proposed idea}
The proposed approach introduces two complementary methods to optimize computational complexity. Firstly, instead of computing CWT using a pre-defined Morlet wavelet kernel, the approach allows for the adjustment of the Morlet wavelet's length. This modification provides flexibility in balancing temporal resolution and computational cost. For applications requiring high temporal resolution, users can increase the wavelet length \textit{(WL)} to capture more detailed temporal features in the signal. Conversely, for applications prioritizing lower computational cost, the wavelet length can be reduced accordingly.

Secondly, instead of generating a complete output sequence with a length equivalent to the input signal for each scale in the CWT, or reducing the number of samples in the original signal \cite{10797444}, the proposed method samples the intermediate output during the transformation process using a predefined hop size \textit{(H)}. This down-sampling is incorporated directly into the convolution step at each scale, thereby reducing the number of extracted temporal features and lowering the computational complexity of the CWT. Furthermore, by performing a grid search over the parameters \textit{(WL)} and \textit{(H)}, users can identify configurations that achieve a favorable trade-off between computational cost and model performance.

\begin{figure}[t]
\centering
\includegraphics[width=0.5\textwidth]{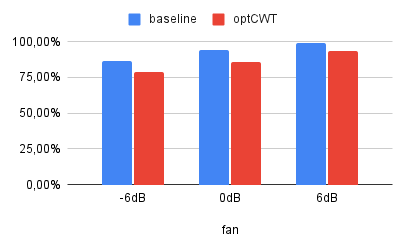}
\caption{Prediction performance AUC-ROC of models on audio of fan}
\label{FanChart}
\end{figure}

\begin{figure}[t]
\centering
\includegraphics[width=0.5\textwidth]{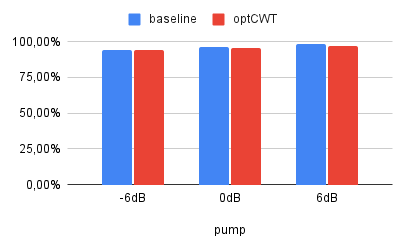}
\caption{Prediction performance AUC-ROC of models on audio of pump}
\label{PumpChart}
\end{figure}

\begin{figure}[t]
\centering
\includegraphics[width=0.5\textwidth]{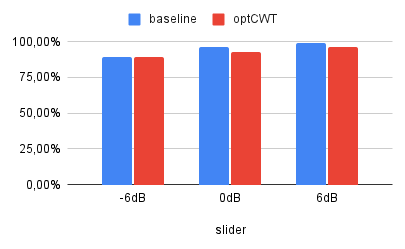}
\caption{Prediction performance of AUC-ROC models on audio of slider}
\label{SliderChart}
\end{figure}

\begin{figure}[t]
\centering
\includegraphics[width=0.5\textwidth]{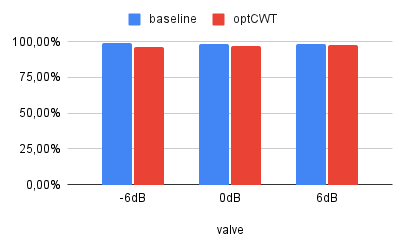}
\caption{Prediction performance AUC-ROC of models on audio of valve}
\label{ValveChart}
\end{figure}

\section{EXPERIMENT}
\subsection{Dataset}

The present study uses the MIMII dataset \cite{purohit2019mimii}, which comprises acoustic recordings collected from industrial environments. The dataset includes sound data from various types of machinery, specifically fans, pumps, sliders, and valves. For each machine type, the dataset categorizes audio recordings into two classes: normal, representing properly functioning machinery, and abnormal, corresponding to malfunctioning machinery. The primary objective of utilizing this dataset is to develop and train models for the detection of machinery faults based on audio signals. To enhance robustness, the recordings are superimposed with environmental background noise at three distinct levels of signal-to-noise ratio (SNR): -6 dB, 0 dB, and 6 dB. The dataset consists of 54,507 audio files, each with a duration of 10 seconds. All recordings are sampled at 16 kHz, yielding 160,000 samples per audio file.

\subsection{Scalogram generation}

The process of scalogram generation is illustrated in Fig. \ref{ScalogramGeneration}, and is herein referred to as optCWT (optimal continuous wavelet transform). Initially, a wavelet kernel of length \textit{WL} is defined at a given scale \textit{S}. By specifying the wavelet length, the transformation can be tailored to balance the trade-off between prediction performance and computational cost. Specifically, longer wavelets may enhance model performance at the expense of increased computational complexity, whereas shorter wavelets may reduce computational demands with potentially lower performance. Subsequently, the discrete audio signal \textit{x(n)} of length \textit{N}, as obtained from the aforementioned dataset, is convolved with the wavelet kernel using the Fast Fourier Transform (FFT), in accordance with the standard wavelet transform procedure. The resulting output is then downsampled by a factor of \textit{H}, rather than retaining its original resolution. This downsampling operation reduces both memory usage and computational overhead, thereby optimizing the efficiency of subsequent processing steps, such as heat maps visualization. The transformation is iteratively performed for the same audio signal across the specified range of wavelet scales. Upon completion, a coefficient matrix of dimensions \textit{(N/H, S)} is produced, which is more compact compared to the standard wavelet transform coefficient matrix of size \textit{(N, S)}. This reduced matrix is subsequently utilized to generate heatmaps, referred to as scalograms. The implementation for scalogram generation in this study has been made publicly available via the GitHub repository referenced in \cite{github:optimal_scalogram}, and a pull request proposing the findings to upgrade the Pywavelets library is already created \cite{github:pull_request}. The configuration of the transform is based on the benchmark study, employing 128 scales of range (2, 129) \cite{10.1007/978-3-031-84457-7_41}. For visual comparison, the scalograms produced by both the conventional CWT and the proposed optCWT are presented side by side in Fig. \ref{Scalograms} (a) and (b) respectively. As observed, the CWT-based scalogram exhibits a more continuous and densely concentrated energy distribution, providing a more detailed representation. In contrast, the optCWT scalogram displays a sparser and more discrete distribution of energy, reflecting the trade-offs introduced by the optimized transformation.

\subsection{Acoustic recognition task}

To perform the acoustic recognition task, a CNNs model is developed and uses the scalogram images generated in the previous step as input data. The model is trained and tested for performance of fault detection via acoustic recognition. The performance of the optCWT-generated scalograms on the CNNs is evaluated and compared to that of scalograms produced by the conventional CWT to assess the efficacy of optCWT in detecting anomalous sounds. The developed pipeline is identical to the benchmark study \cite{10.1007/978-3-031-84457-7_41} for comparison purpose. The settings for scales, heat map generation, and CNN parameters are also adopted from the referenced study. These include an input image size of 512×512 pixels, convolutional layers with a kernel size of 3×3, and a learning rate of 0.001. The model is trained using a batch size of 64 for 32 epochs. The only extra steps are the definition of wavelet kernel length and sampling the output of CWT. Two main parameters of this study, wavelet length \textit{WL} and hop size \textit{H} are examined in grid search manner to find an optimal set \textit{WL = 64}, \textit{H = 128}, which is significantly reduce the computation cost while achieving a satisfactory performance.

The research employs the PyWavelets library for wavelet transformation, the TensorFlow library for implementing the binary classification CNNs model, and the Area Under the Curve of the Receiver Operating Characteristic (AUC-ROC) \cite{provost1998case} as the prediction performance metric.

\subsection{Results}

\begin{table}[]
\centering
\caption{Performance of models for audio of fan}
\label{tab:fan}
\begin{tabular}{|c|c|c|c|}
\hline
               & \textbf{Baseline} & \textbf{optCWT} \\ \hline
\textbf{-6 dB} & 86,71\%           & 78,73\%            \\ \hline
\textbf{0 dB}  & 94,41\%           & 85,98\%             \\ \hline
\textbf{6 dB}  & 99,24\%           & 93,62\%             \\ \hline
\end{tabular}
\end{table}

\begin{table}[]
\centering
\caption{Performance of models for audio of pump}
\label{tab:pump}
\begin{tabular}{|c|c|c|c|}
\hline
               & \textbf{Baseline} & \textbf{optCWT} \\ \hline
\textbf{-6 dB} & 93,91\%           & 93,96\%            \\ \hline
\textbf{0 dB}  & 96,21\%           & 95,34\%             \\ \hline
\textbf{6 dB}  & 98,62\%           & 97,13\%             \\ \hline
\end{tabular}
\end{table}

\begin{table}[]
\centering
\caption{Performance of models for audio of slider}
\label{tab:slider}
\begin{tabular}{|c|c|c|c|}
\hline
               & \textbf{Baseline} & \textbf{optCWT} \\ \hline
\textbf{-6 dB} & 89,03\%           & 89,40\%            \\ \hline
\textbf{0 dB}  & 96,44\%           & 92,67\%             \\ \hline
\textbf{6 dB}  & 98,85\%           & 96,24\%             \\ \hline
\end{tabular}
\end{table}

\begin{table}[]
\centering
\caption{Performance of models for audio of valve}
\label{tab:valve}
\begin{tabular}{|c|c|c|c|}
\hline
               & \textbf{Baseline} & \textbf{optCWT} \\ \hline
\textbf{-6 dB} & 98,92\%           & 96,64\%            \\ \hline
\textbf{0 dB}  & 98,61\%           & 96,87\%             \\ \hline
\textbf{6 dB}  & 98,76\%           & 97,54\%             \\ \hline
\end{tabular}
\end{table}

The prediction performance of the models across various audio types is documented in Tables \ref{tab:fan}, \ref{tab:pump}, \ref{tab:slider} and \ref{tab:valve}, and visualized in Figures \ref{FanChart}, \ref{PumpChart}, \ref{SliderChart} and \ref{ValveChart}. As observed, the predictive performance of both models improves with increasing SNR levels across all four types of machines. This improvement can be attributed to the fact that higher SNR levels yield more distinguishable features, thereby facilitating more effective training and prediction. In comparing the two models, the baseline model consistently demonstrates performance that is comparable to or superior to that of optCWT, which is likely due to the higher resolution of its wavelet transform.

\begin{table}[]
\centering
\caption{Computational load for a single file}
\label{tab:ComputationalComplexity}
\begin{tabular}{|c|c|c|}
\hline
\textbf{Time}       & \textbf{Single audio} & \textbf{Entire dataset} \\ \hline
\textbf{Baseline} & 8.09s                 & 122.5hrs                \\ \hline
\textbf{optCWT}   & 1.15s                 & 17.5hrs                 \\ \hline
\end{tabular}
\end{table}

\begin{figure}[t]
\centering
\includegraphics[width=0.5\textwidth]{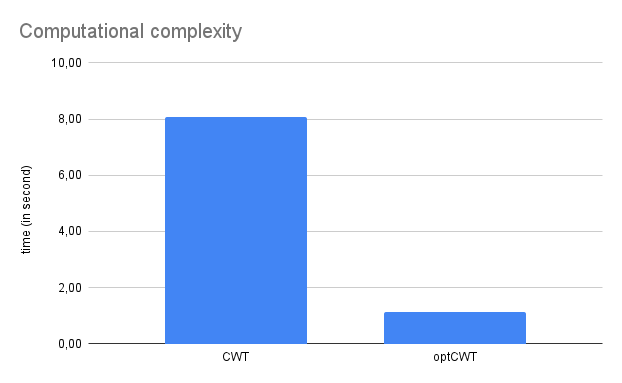}
\caption{Computational complexity in generation of a single file in second}
\label{ComputationalComplexity}
\end{figure}

Notably, the results for the fan and valve machines warrant further discussion. For the fan, both models exhibit lower performance relative to the other machines, with a larger performance gap between the baseline and optCWT. A possible explanation is the stationary nature of fan audio signals \cite{purohit2019mimii}, which are less effectively represented by wavelet transforms. As a result, the baseline model, which employs a higher-resolution wavelet transform, performs better than optCWT, which utilizes a lower-resolution transform. Conversely, in the case of the valve, which produces non-stationary audio signals \cite{purohit2019mimii}, both models achieve stable and superior performance across all SNR levels, with a smaller performance gap. This can be attributed to the multi-resolution analysis capability of the wavelet transform, which is particularly effective for extracting features from non-stationary signals. This leads to enhanced performance for both models and reduces the disparity between them. These findings are consistent with the results reported in previous studies \cite{10.1007/978-3-031-84457-7_41}, \cite{10797444}.

Overall, optCWT delivers performance that is fair and comparable to that of the baseline model, indicating its potential applicability in practical scenarios. Moreover, optCWT offers a substantial computational advantage as recorded in Table \ref{tab:ComputationalComplexity} and illustrated in Fig. \ref{ComputationalComplexity}. In the experimental setup, the generation of optCWT for a single file required only 1.15 seconds, whereas the baseline's CWT required 8.09 seconds - approximately seven times longer. For the complete dataset comprising 54,507 files, the total processing time amounted to 17.5 hours for optCWT, compared to 122.5 hours for the baseline CWT. These results underscore the significant computational efficiency of optCWT.

\section{CONCLUSION AND DISCUSSION}

This research has developed an efficient method for acoustic recognition by accepting a minor reduction in prediction performance in exchange for a substantial decrease in computational complexity. This trade-off is particularly advantageous for applications requiring real-time processing or operating under limited computational resources.

Future research should explore the use of alternative wavelet types, such as the Mexican Hat wavelet and the Shannon wavelet, to establish a set of wavelets compatible with the proposed method for further reducing computational complexity. Additionally, evaluating the method on diverse datasets is anticipated to enhance the generalizability of its applicability across various types of audio data.

\bibliographystyle{IEEEtran}
\IEEEtriggeratref{34}
\bibliography{CWTH}
\vspace{12pt}
\end{document}